\documentclass{emulateapj}

\newcommand{\unit}[1]{\,{\rm #1}}

\begin{document}

\title{Effect of the Coriolis Force on the Hydrodynamics of Colliding Wind Binaries}

\author{M. Nicole Lemaster, James M. Stone, and Thomas A. Gardiner\altaffilmark{1}}
\affil{Department of Astrophysical Sciences, Princeton University, Princeton, NJ 08544}
\altaffiltext{1}{Present address: 3915 Rayado Pl NW, Albuquerque, NM 87114}

\shorttitle{Hydrodynamics of CWBs}
\shortauthors{Lemaster, Stone, \& Gardiner}

\begin{abstract}

Using fully three-dimensional hydrodynamic simulations, we investigate
the effect of the Coriolis force on the hydrodynamic and observable
properties of colliding wind binary systems.  To make the calculations
tractable, we assume adiabatic, constant velocity winds.  The neglect
of radiative driving, gravitational deceleration, and cooling limit
the application of our models to real systems.  However, these
assumptions allow us to isolate the effect of the Coriolis force,
and by simplifying the calculations, allow us to use a higher
resolution (up to $640^3$) and to conduct a larger survey of parameter
space.  We study the dynamics of collidng winds with equal mass
loss rates and velocities emanating from equal mass stars on circular
orbits, with a range of values for the ratio of the wind to orbital
velocity.  We also study the dynamics of winds from stars on
elliptical orbits and with unequal strength winds.  Orbital motion
of the stars sweeps the shocked wind gas into an Archimedean spiral,
with asymmetric shock strengths and therefore unequal postshock
temperatures and densities in the leading and trailing edges of the
spiral.  We observe the Kelvin-Helmholtz instability at the contact
surface between the shocked winds in systems with orbital motion
even when the winds are identical.  The change in shock strengths
caused by orbital motion increases the volume of X-ray emitting
post-shock gas with $T > 0.59 \unit{keV}$ by $63\%$ for a typical
system as the ratio of wind velocity to orbital velocity decreases
to $V_w/V_o = 2.5$.  This causes increased free-free emission from
systems with shorter orbital periods and an altered time-dependence
of the wind attenuation.  We comment on the importance of the effects
of orbital motion on the observable properties of colliding wind
binaries.

\end{abstract}

\keywords{binaries: general --- hydrodynamics --- stars: early-type --- stars: winds, outflows --- stars: Wolf-Rayet --- X-rays: stars}

\section{INTRODUCTION}

A large proportion (at least 39\%) of Wolf-Rayet (WR) stars in the solar
neighborhood have been observationally confirmed to lie in binary systems
(van der Hucht 2001).  These stars have dense, highly-supersonic,
radiatively-driven winds that, upon colliding with the wind from a companion,
produce strong shocks that compress and heat the gas to temperatures
high enough to produce substantial X-ray flux.  These colliding wind binaries
(CWBs), which can contain a pair of WR stars, a WR star with an OB supergiant
companion, or a pair of OB stars, were first proposed by Prilutskii \& Usov
(1976) and Cherepashchuk (1976).  This X-ray excess has since been 
observationally confirmed (Pollock 1987).

For more than a decade, the hydrodynamics of CWBs have been investigated
using two-dimensional (axisymmertric) hydrodynamical simulations (e.g. Luo,
McCray, \& Mac Low 1990; Stevens, Blondin, \& Pollock 1992) in order to compute
synthetic X-ray spectra for comparison to observations.  A variety of physical
effects that are important for the dynamics of the winds have been identified,
including radiative inhibition (Stevens \& Pollock 1994) and sudden radiative
braking (Gayley, Owocki, \& Cranmer 1997).  The former is the effect of the
radiation field of one star decreasing the radiative acceleration of the wind
from the other, while the latter is where the radiation field of one star is
intense enough to actually cause a net deceleration of the other star's wind.
While including such effects leads to more realistic models of CWBs
(e.g. Henley, Stevens, \& Pittard 2005), most current simulations
are typically still done in only two-dimensions, assuming axisymmetry
around the line joining the two stars.  This requires neglecting the
Coriolis force resulting from orbital motion.  Including the Coriolis
force requires fully three-dimensional hydrodynamical models which,
up to now, have been largely untractable.  

In this paper, we investigate the effect of orbital motion on the
hydrodynamic and observable properties of the wind collision region by
performing fully three-dimensional hydrodynamic simulations of CWBs.
We conduct a parameter survey in orbital velocity for stars on circular
orbits, and also present models for stars on elliptical orbits and with
unequal strength winds.

Our focus in this paper is the three-dimensional dynamics of the wind
collision region, thus to simplify the problem we neglect a number of
physical processes that can be important in real systems, but would serve to
complicate the interpretation of our results.  In particular, we neglect
radiative cooling of the postshock gas and do not try to model the radiative
driving of the winds.  Instead of explicitly including gravity in our
simulations, we assume that gravitational deceleration balances
radiative acceleration in the unshocked winds and give the winds a constant
velocity profile.  This velocity is chosen to be the unshocked wind velocity
given by a beta velocity law at the distance of the stagnation point (see
eq. [\ref{eq:beta}] for more details).  We discuss in \S\ref{sec:conclusion}
possible implications of neglecting gravitational force on the unshocked wind
of the companion star as well as the post-shock gas.  Since we do not model
radiative effects, there will be no sudden radiative braking or radiative
inhibition in our simulations.  Sudden radiative braking is thought to occur
only with strongly unequal winds (Gayley et al. 1997), which we will not
consider here.

While for most early-type binaries it is important to consider radiative 
cooling of the postshock gas by line emission, we leave this to future 
research.  We have chosen a sample system that could be composed of a pair
of O-stars, which we find is neither purely adiabatic nor strongly radiative
(see \S\ref{sec:systems}) using the appropriate formula in Antokhin, Owocki,
\& Brown (2004).  We will comment on how including more realistic radiative
effects on the winds might change our results.

It is important to note ours are not the first fully three-dimensional
simulations of CWBs; Walder (1998) showed that orbital motion can produce
Archimedian spirals similar to those visible in infrared images of WR
98a (Monnier, Tuthill, \& Danchi 1999) and WR 104 (Tuthill, Monnier,
\& Danchi 1999).

The organization of this paper is as follows.  In \S\ref{sec:numerical}
we describe the details of our simulations and the diagnostics we use
to interpret our results. In \S\ref{sec:results} we present analysis
of the post-shock temperatures, emissivity, column density, and
instabilities that result for colliding winds from identical stars on
circular orbits, with a range of values for the ratio of wind velocity
to orbital velocity.  We also consider systems with eccentric orbits
and unequal winds.  Finally, in \S\ref{sec:conclusion} we discuss the
implications by applying our results to WR 20a.

\section{METHOD}\label{sec:numerical}

We present three-dimensional hydrodynamic simulations conducted on a cartesian grid with the Athena code (Gardiner \& Stone 2005, 2006).
We solve the equations of hydrodynamics
\begin{equation}
\frac{\partial \rho}{\partial t} + {\bf \nabla} \cdot (\rho {\bf v}) = 0,
\end{equation}
\begin{equation}
\frac{\partial \rho {\bf v}}{\partial t} + {\bf \nabla} \cdot (\rho {\bf vv} + P) = 0,
\end{equation}
and
\begin{equation}
\frac{\partial E}{\partial t} + {\bf \nabla} \cdot \big[ (E + P) {\bf v} \big] = 0,
\end{equation}
where $E$ is the total energy density,
\begin{equation}
E \equiv \epsilon + \frac{1}{2} \rho {\bf v v},
\end{equation}
$\epsilon$ is the internal energy density, and we adopt an ideal gas
equation of state, $P = (\gamma - 1) \epsilon$, where $\gamma = 5/3$
is the ratio of specific heats.  We use the Roe Riemann solver augmented
with the H-correction of Sanders, Morano \& Druguet (1998) to prevent the
carbuncle phenomenon (Quirk 1994).  We use outflow boundary conditions
and run our simulations for up to three orbital periods.  Most of our
simulations were run on $256^3$ grids with boxes of length $L = 2.5 \alpha 
\unit{AU}$ (where the seperation between the stars is $\alpha 
\unit{AU}$), however we also ran three simulations on $640^3$ grids, two of 
them with $L = 6.25 \alpha \unit{AU}$ and the third with $L = 5 \alpha
\unit{AU}$.
We will define the parameter $\alpha$ in the next section.

\subsection{Initial Conditions}\label{sec:windsol}

Our stars are given a steady, spherically-symmetric, radial wind by forcing the correct density, pressure, and velocity profiles onto a small region of the grid.
The density and pressure of the wind are given by
\begin{equation}\label{eq:denprofile}
\rho (r) = \rho_0 \Big( \frac{r_0}{r} \Big)^2
\end{equation}
and
\begin{equation}\label{eq:prprofile}
P(r) = P_0 \Big( \frac{r_0}{r} \Big)^{10/3}.
\end{equation}
Our wind is generated with a constant velocity, $v_w$, taken to be the velocity
at the stagnation point in the circular-orbit case.
The constants $\rho_0$ and $P_0$ are defined to give the desired mass loss rate, $\dot{M}$, and wind mach number, $\mathcal{M}_0$, at the edge of the mask, $r_0$, using
\begin{equation}\label{eq:massloss}
\dot{M} = 4\pi r_0^2 \rho_0 v_w
\end{equation}
and
\begin{equation}\label{eq:windmach}
\mathcal{M}_0 = v_w \sqrt{\frac{\rho_0}{\gamma P_0}},
\end{equation}
where $r_0$ is some fiducial radius, taken to be $r_0 = 0.195 \alpha \unit{AU}$.
The density, pressure, and velocity in the simulations are in an arbitrary
system of units.

The wind and orbital parameters in our simulations can been parameterized by
the variables $\alpha$, $\nu$, and $\eta$, defined in Table \ref{tab:pars1}.
These quanities, which are generally of order unity for early-type binary
systems, can be varied in order to apply our results to a variety of systems.

\begin{deluxetable}{ll}
\tablecolumns{2}
\tablewidth{3in}
\tablecaption{Simulation Parameters\label{tab:pars1}}
\tablehead{
\colhead{Parameter} & \colhead{Scaling Used}}
\startdata
Semi-major axis of system & $\alpha = a / 1 \unit{AU}$ \\
Wind velocity & $\nu = v_w / 1000 \unit{km} \unit{s^{-1}}$ \\
Mass loss rate & $\eta = \dot{M} / 10^{-5} \unit{M_\odot} \unit{yr^{-1}}$ \\
Wind mach number & $\mathcal{M}_0 = 30$
\enddata
\end{deluxetable}

Initially, the entire computational domain is filled with a stationary ambient medium with density $\rho_{amb} = 1.09 \times 10^{-17} \eta \alpha^{-2} \nu^{-1} \unit{g} \unit{cm^{-3}}$ and sound speed $c_{s,amb} = 21.0 \nu \unit{km} \unit{s^{-1}}$.
The masks are then initialized with a radius $2 r_0$.
The properties of the ambient medium were chosen for numerical, not physical, reasons.
We evolve the system for much longer than a wind-crossing time to make sure the ambient medium has been completely driven off the grid before starting our analysis.
See the appendix for the details of how we generate our wind.

\subsection{Systems Investigated}\label{sec:systems}

To investigate the effect of orbital motion on both the hydrodynamic and observable properties of the collision region, we take a binary system with a circular orbit and vary the ratio of the wind velocity to orbital velocity.
The mass of the stars and the orbital period are varied so as to keep the semi-major axis of the orbit constant while varying the orbital velocity.

The sample system to which we will apply our results has $\alpha = 0.69$,
$\nu = 0.85$, and $\eta = 0.065$.
This gives $a = 0.69 \unit{AU}$, $\dot{M} = 6.5 \times 10^{-7} \unit{M_\odot} \unit{yr^{-1}}$, and $v_w = 850 \unit{km/s}$ at the stagnation point.
Given a beta velocity law,
\begin{equation}\label{eq:beta}
v(r) = v_\infty \Big( 1-\frac{R_*}{r} \Big)^\beta,
\end{equation}
with $\beta = 1$, we find $v_\infty \approx 1160 \unit{km} {s^{-1}}$ for its terminal velocity, assuming $R_* = 20 R_\odot$.
This set of parameters gives for each of the stellar masses
\begin{eqnarray}
M_* &=& 2.25 \times 10^3 \alpha \nu^2 \Big( \frac{V_w}{V_o} \Big)^{-2} \unit{M_\odot} \\
&=& 1.12 \times 10^3 \Big( \frac{V_w}{V_o} \Big)^{-2} \unit{M_\odot}.
\end{eqnarray}
It should be noted that stars with masses above those given by $V_w/V_o \approx 3.6$ for this sample system have not yet been detected.
Using the method of Antokhin et al (2004) to determine if a system with these parameters is radiative ($\alpha < \alpha_{rad}$) or adiabatic ($\alpha > \alpha_{rad}$), we find
\begin{equation}\label{eq:alpharad}
\alpha_{rad} = 4.66 \frac{\eta}{\nu^5} = 0.683,
\end{equation}
meaning that our system is very slightly on the adiabatic side.
Despite neglecting radiative effects in our simulations, we feel there is still something to be learned from studying this hypothetical set of systems.

We start by presenting the results of circular-orbit simulations with $V_w/V_o = \infty$ and $V_w/V_o = 2.5$ in a large box ($L = 6.25 \alpha \unit{AU}$) on a $640^3$ grid.
We then compare a set of simulations with varying $V_w/V_o$, whose names and velocity ratios are given in Table \ref{tab:pars2}, in a smaller box ($L = 2.5 \alpha \unit{AU}$) on a $256^3$ grid.
This preserves cell size between the large- and small-box runs.

\begin{deluxetable}{cc}
\tablecolumns{2}
\tablewidth{1in}
\tablecaption{Circular-orbit Simulation Parameters\label{tab:pars2}}
\tablehead{
\colhead{Name} & \colhead{$V_w/V_o$}}
\startdata
$S$ & $\infty$ \\
$C10$ & 10.0 \\
$C5$ & 5.0 \\
$C3.5$ & 3.5 \\
$C2.5$ & 2.5
\enddata
\end{deluxetable}

We compare two additional simulations with eccentric orbits to our corresponding circular-orbit simulations.
Simulations $SE2.5$ and $E2.5$, respectively, have identical properties to simulations $S$ and $C2.5$, except with eccentricity $e = 0.2$.
In simulation $SE2.5$, the separation of the stars is varied over time with zero orbital velocity perpendicular to the line of centers, as was done by Pittard (1998).

Finally, we consider simulations similar to $C2.5$ but with unequal winds.
Simulation $CM2.5$ has identical parameters to $C2.5$, except that, for the secondary (on the negative x-axis when the simulation is initialized), $\eta_2 = (2/3)\eta$, giving a lower mass-loss rate. 
Simulation $CW2.5$ also has identical parameters to $C2.5$, except that, for the secondary, $\nu_2 = (2/3)\nu$, giving a lower wind velocity.
The modified wind velocity and mass-loss rate were chosen so as to give a wind momentum ratio of 1.5 in all unequal wind simulations.

\subsection{Diagnostics}\label{sec:diagnostics}

We analyze our simulations by comparing post-shock temperatures, free-free
emission from post-shock gas above a minimum temperature, and column density
to key locations in the system.
So that our results can be scaled to other systems, we leave most of our
results parameterized in terms of $\alpha$, $\nu$, and $\eta$, described in
\S\ref{sec:windsol}, as well as $\bar{\mu}$, the mean particle mass in amu,
and $Z$, the charge of the ion, but substitute for $\mathcal{M}_0$.
Stevens et al. (1992) give $\bar{\mu} = 0.6$ for solar abundances, $\bar{\mu} = 1.3$ for WN stars, and $\bar{\mu} = 1.4$ for WC stars, assuming full ionization.

Due to the scalings chosen for our simulations, one can convert the densities from our simulations to physical units using
\begin{equation}
\rho = 1.09 \times 10^{-16} \eta \alpha^{-2} \nu^{-1} \widetilde{\rho} \unit{g} \unit{cm^{-3}},
\end{equation}
and the temperatures by
\begin{equation}
T = 0.0106 \bar{\mu} \nu^2 \frac{\widetilde{P}}{\widetilde{\rho}} \unit{keV},
\end{equation}
where $\widetilde{P}$ and $\widetilde{\rho}$ are the pressure and density, respectively, in simulation units.

We calculate the power radiated per unit volume by free-free emission over all frequencies and directions using
\begin{equation}
\Lambda_{ff} = 2.1 \times 10^{-8} Z^2 \bar{\mu}^{-2} T_{\unit{keV}}^{1/2} \rho_{-16}^2 \unit{erg} \unit{s^{-1}} \unit{cm^{-3}},
\end{equation}
where $\rho_{-16}$ is the density in units of $10^{-16} \unit{g} \unit{cm^{-3}}$.
We have assumed the frequency-averaged Gaunt factor to be $\langle g_{ff}
\rangle = 1.2$ with the understanding that this will only give an accuracy of
about $20\%$ (Rybicki \& Lightman, 1979).
We integrate this equation over the post-shock volume to find the total power radiated by free-free emission, $\mathcal{P}_{ff}$.
The kinetic luminosity of each wind is $L_W = 3.15 \times 10^{36} \eta \nu^2 \unit{erg} \unit{s^{-1}}$.

To see how the light curve would be attenuated by the wind, we calculate the column density to each star and to the center of mass of the system, the source of the hardest X-ray emission from the post-shock gas.
We use lines of sight inclined 25$^\circ$ from the normal to the orbital plane, whose projections onto the orbital plane are aligned with the zero-phase line of centers.
Integrating from the edge of the box to either the center of mass or the edge of the stellar mask, we find for the column density
\begin{equation}
N = 8.95 \times 10^{20} \bar{\mu}^{-1} \int \rho_{-16} dr_{\rm AU} \unit{cm^{-2}}.
\end{equation}
Since we are only integrating from the edge of the box, we will miss the contribution to the column density from the edge of the box to the observer.
The contribution from a spherically-symmetric, unshocked wind outside the box, along the line of sight to one of the stars, would be
\begin{equation}
\Delta N = 3.64 \times 10^{22} \frac{\eta}{\bar{\mu} \nu} L_{\rm AU}^{-1} \unit{cm^{-2}}.
\end{equation}
For our sample system in a small box ($L = 2.5 \alpha \unit{AU}$), this gives $\Delta N = 1.62 \times 10^{21} \bar{\mu}^{-1} \unit{cm^{-2}}$.
In the largest box ($L = 6.25 \alpha \unit{AU}$), we find $\Delta N = 6.46 \times 10^{20} \bar{\mu}^{-1} \unit{cm^{-2}}$.

\section{RESULTS}\label{sec:results}

\subsection{Identical Winds on Circular Orbits}

\begin{figure}
\epsscale{0.3}
\plotone{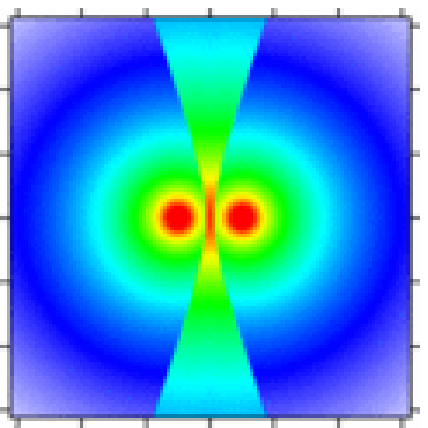}
\plotone{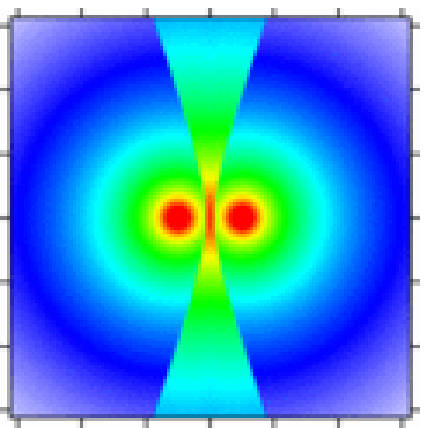}
\plotone{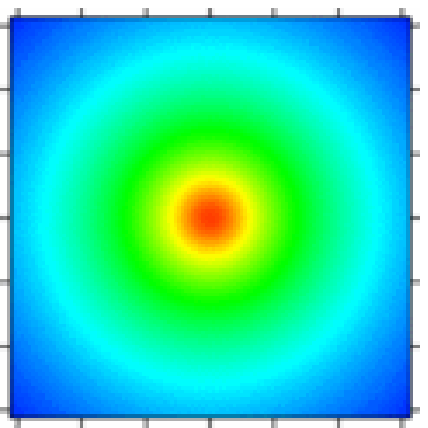} \\
\plotone{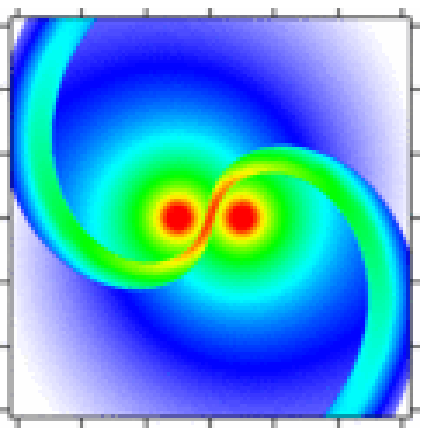}
\plotone{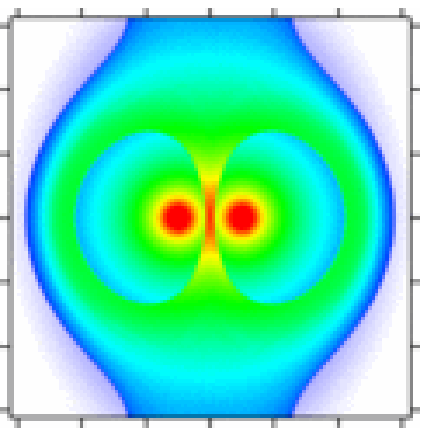}
\plotone{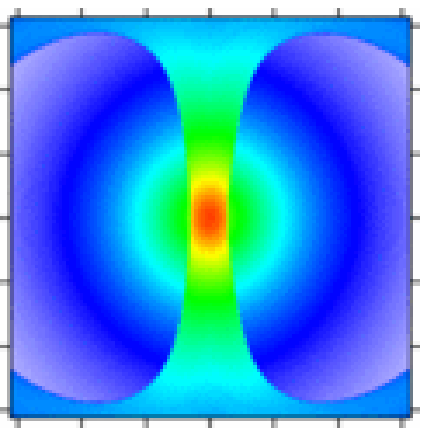}
\figcaption{
Density in large-box simulation $S$
(\emph{a}) in orbital plane, 
(\emph{b}) through the stars normal to the orbital plane, and
(\emph{c}) normal to the previous two slices.
Density in large-box simulation $C2.5$
(\emph{d}) in orbital plane, 
(\emph{e}) through the stars normal to the orbital plane, and
(\emph{f}) normal to the previous two slices.
The color scale is logarithmic, with $\alpha^2 \nu \eta_{-1} \rho_{-16}$ from 1 (white) to 700 (red).
The tick marks are spaced $\alpha \unit{AU}$.
\label{fig:den640}}
\end{figure}

\begin{figure}
\epsscale{0.3}
\plotone{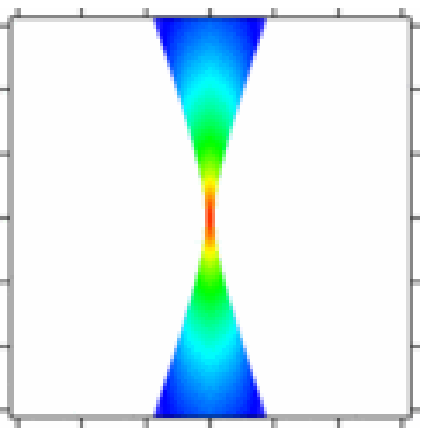}
\plotone{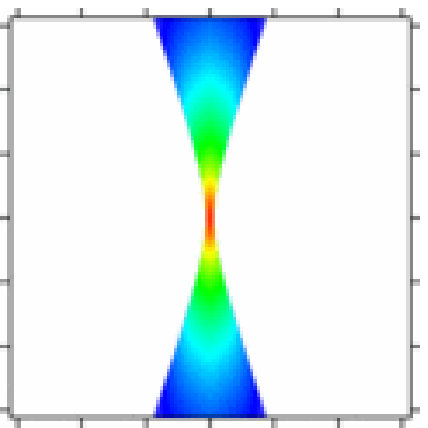}
\plotone{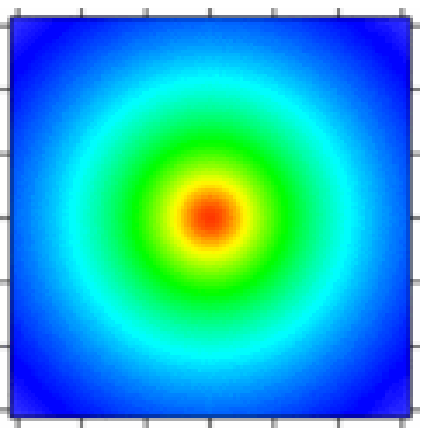} \\
\plotone{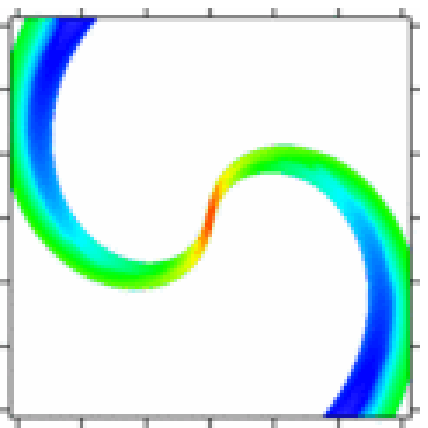}
\plotone{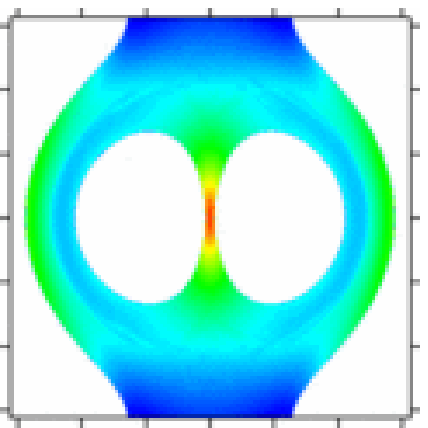}
\plotone{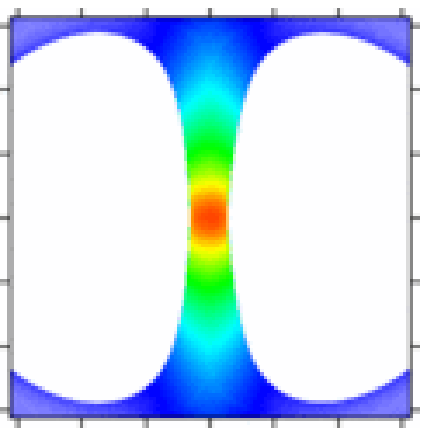}
\figcaption{
Temperature in large-box simulation $S$
(\emph{a}) in the orbital plane, 
(\emph{b}) through the stars normal to the orbital plane, and
(\emph{c}) normal to the previous two slices.
Temperature in large-box simulation $C2.5$
(\emph{d}) in the orbital plane, 
(\emph{e}) through the stars normal to the orbital plane, and
(\emph{f}) normal to the previous two slices.
The color scale is logarithmic, with $\bar{\mu}^{-1} \nu^{-2} T_{\unit{keV}}$ from 0.05 (white) to 2.5 (red).
The tick marks are spaced $\alpha \unit{AU}$.
\label{fig:temper640}}
\end{figure}

\begin{figure}
\epsscale{1.0}
\plotone{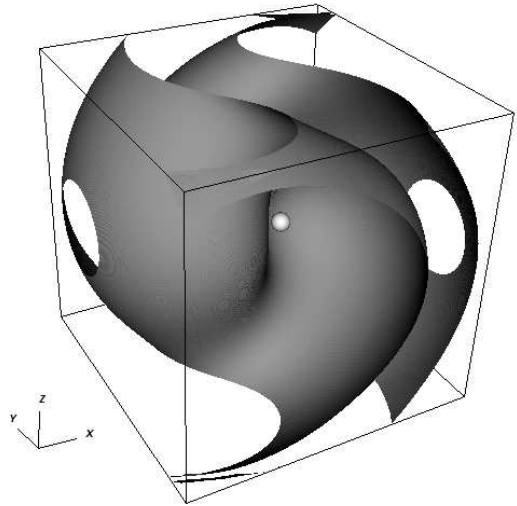}
\figcaption{
Shock structure for large-box circular-orbit simulation with $V_w/V_o = 2.5$.
The surface shown is $T = 0.06125 \bar{\mu} \nu^2 \unit{keV}$.
Small spheres mark the size and location of the stellar masks.
The stars are moving counter-clockwise in the $x$-$y$ plane.
\label{fig:shock-640d}}
\end{figure}

Figure \ref{fig:den640} shows the density for the large-box ($L = 6.25 \alpha \unit{AU}$)
runs of $S$ and $C2.5$ in three orthogonal cut planes through the center
of mass of the system.  Figure \ref{fig:temper640} shows the temperature
in the same three planes.  The shape of the shock front for $C2.5$, with
$V_w/V_o = 2.5$, differs substantially from that for the stationary star
case, $S$.  The shocks wrap around the stars, breaking the axisymmetry that
is present when the stars are held stationary.  The absence of shocked
material in the two circular regions surrounding the stars in the slices
normal to the orbital plane is a result of the curvature of the shock.
The shocked material that would be projected onto those regions is on
the near side of the left star and on the far side of the right star,
relative to the viewer, as is apparent from the slice in the orbital
plane.  The three-dimensional structure of the post-shock region is more
visible in Figure \ref{fig:shock-640d}, which traces the shock fronts using
a surface of constant temperature.  Small spheres mark the size and position
of the stellar masks.  There is $63\%$ more volume at $T > 0.63 \bar{\mu}
\nu^2 \unit{keV}$ in $C2.5$ than in $S$ but only $20\%$ more at $T > 0.875
\bar{\mu} \nu^2 \unit{keV}$.

\begin{figure}
\epsscale{1.0}
\plotone{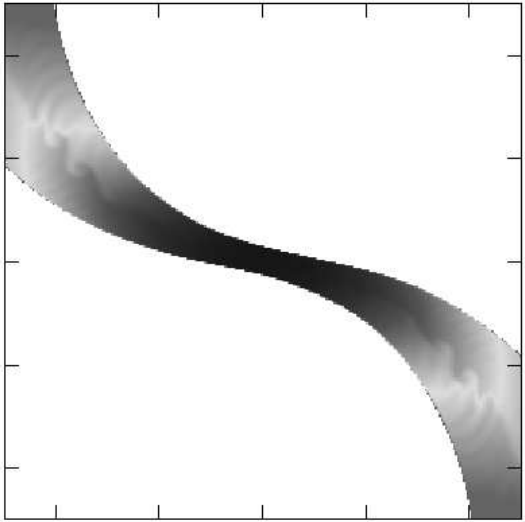}
\figcaption{
Temperature in the orbital plane of small-box simulation $C5$.
The image has been enhanced to make the effect of the Kelvin-Helmholtz instability on the contact surface more easily visible.
The tick marks are spaced $0.5 \alpha \unit{AU}$.
\label{fig:kh-temper}}
\end{figure}

The maximum post-shock density and temperature are not significantly
different between $C2.5$ and $S$ when the magnitude of the time-variation
in $C2.5$ is considered.
The temperatures on the leading side of the shock in $C2.5$ are higher
than on the trailing side;  conversely, the densities
are lower.  The thickness of the post-shock
region between the stars is the same for both $C2.5$ and $S$.  In both
cases, the post-shock region is centered on the center of mass of the
system.  The numerical methods used in the Athena code keep the shocks
very thin, leading to a stair-stepping of the density and temperature
as the curved shock crosses the Cartesian grid.  This stair-stepping
is clearly visible in Figure \ref{fig:kh-temper}, a slice through the
orbital plane of the temperature in the small-box simulation $C5$.

While there is no evidence of the Kelvin-Helmholtz (K-H) instability at
the contact surface bewteen the shocked winds in $S$ (where the stars are
held stationary and the wind velocities are identical), K-H instability is
clearly visible, as in Figure \ref{fig:kh-temper}, in simulations where the
stars are orbiting, even though the winds are identical.  Although over
most of the stellar orbits the effect of the K-H instability on the postshock
gas is very small, intermittently large rolls form which can affect observable
quantities.  For example, the power emitted from gas with $T > 0.875 \bar{\mu}
\nu^2 \unit{keV}$ varies by $\sim 2\%$ as a large roll is advected off the
grid in the small-box simulation $C5$.  The instability is seeded by grid noise
associated with the representation of thin shocks on the Cartesian grid (the
stair-stepping observed in Figure \ref{fig:kh-temper}).  The amplitude of
the grid noise is correlated with the phase of the orbit (direction of the
shock with respect to the grid), therefore we see periodicity in the K-H
rolls at one-quarter of an orbit.  Although the K-H rolls in our
calculations are seeded by grid noise, the effect is real, and could be
seeded by variability or clumpiness in the winds.  The stair-stepping also
introduces grid noise into the volume occuppied by gas at various
temperatures.  For this reason, we try to compare the results for rapidly
orbiting stars to model C10 rather than model S as much as possible in our
parameter survey, since in C10 the motion of the shocks over the grid
averages out the grid noise, yet the orbital motion is small enough to not
significantly affect the structure compared to the stationary case S.

The power emitted by free-free emission from
gas with $T > 0.63 \bar{\mu} \nu^2 \unit{keV}$ is $\mathcal{P}_{ff} = 1.9
\times 10^{36} Z^2 \bar{\mu}^{-3/2} \eta^2 \alpha^{-1} \nu^{-1} \unit{erg}
\unit{s^{-1}}$ for $C2.5$.  This gives the ratio of the free-free power to
the wind luminosity $0.29 Z^2 \bar{\mu}^{-3/2} \eta^2 \alpha^{-1} \nu^{-1}$.
For our binary O-star system, this evaluates to
$2.1 \times 10^{-3} Z^2 \bar{\mu}^{-3/2}$, which
supports not incorporating cooling into our simulations for this particular
system.  The time-average of the power radiated by free-free emission from gas
with $T > 0.63 \bar{\mu} \nu^2 \unit{keV}$ is $13\%$ higher for $C2.5$ than for
$S$.  The standard deviation is $0.24\%$ of the mean for the power emitted from
$C2.5$.

\begin{figure}
\epsscale{1.0}
\plotone{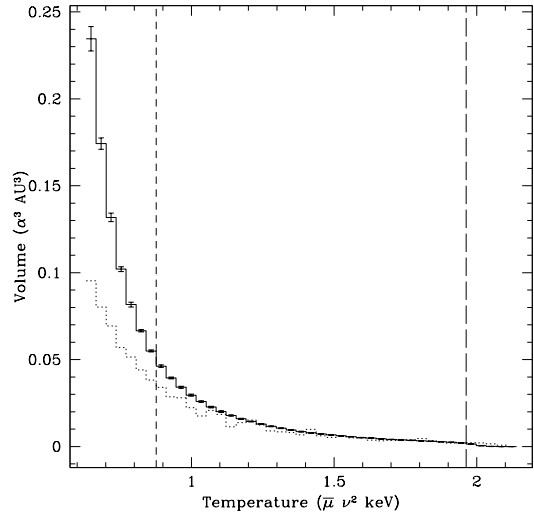}
\figcaption{
Histogram of volume within post-shock temperature bins of width $\Delta T = 0.0212 \bar{\mu} \nu^2 \unit{keV}$, averaged over one orbit, from large-box runs of $C2.5$ (solid) and $S$ (dotted).
The vertical short dash line marks $T = 0.875 \bar{\mu} \nu^2 \unit{keV}$, the temperature cutoff that will be used for the circular-orbit parameter survey analysis.
The vertical long dash line marks $T = 1.96 \bar{\mu} \nu^2 \unit{keV}$, the expected maximum temperature calculated analytically.
\label{fig:highd-temp}}
\end{figure}

Figure \ref{fig:highd-temp} shows volume as a function of post-shock
temperature averaged over one orbital period.  Post-shock temperature
decreases rapidly with distance from the line of centers, leaving
most of the post-shock volume at low temperatures.  Error bars, resulting
from both grid effects and the K-H instability, are shown but are barely
visible except at the lowest temperatures plotted.  As instabilities are
seeded near the center of mass of the system, the volume at the highest
temperatures is affected first.  A wave with an amplitude of a few percent
passes through the temperature distribution, moving from high to low
temperatures, as the gas moves away from the line of centers and off the grid.

As expected, the highest post-shock temperatures are near the line
of centers.  Using the density and pressure profiles described in
\S\ref{sec:windsol}, we derive an unshocked temperature and mach number at
the stagnation point in the stationary-star simulation that would yield a
maximum post-shock temperature of $T = 1.96 \bar{\mu} \nu^2 \unit{keV}$.  This
temperature is marked by the long dashed vertical line in Figure
\ref{fig:highd-temp}.  The high-temperature tail of the distribution extends
beyond this predicted maximum, likely due to multi-dimensional effects not
taken into account in our estimate.  To prevent our results from having a
dependence on orbital phase due solely to the cubical shape of the
computational domain, we choose to consider in our analysis only cells with
post-shock temperatures $T > 0.63 \bar{\mu} \nu^2 \unit{keV}$ for the large-box
runs and $T > 0.875 \bar{\mu} \nu^2 \unit{keV}$ (marked with the short dashed
vertical line in Figure \ref{fig:highd-temp}) for the small-box runs.

\begin{figure}
\epsscale{1.0}
\plotone{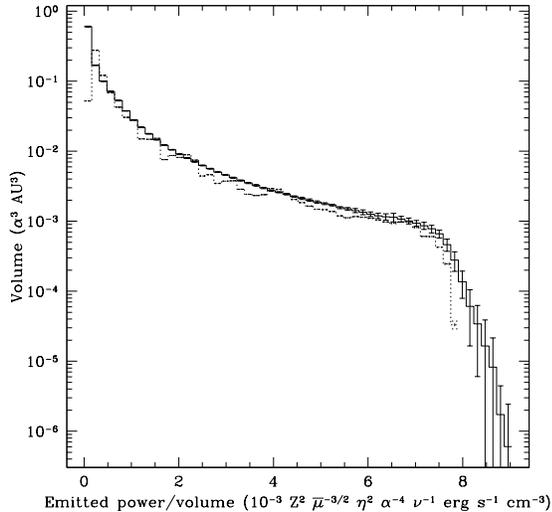}
\figcaption{
Histogram of volume within bins of width
$\Delta \Lambda_{ff} = 1.25 \times 10^{-4} Z^2 \bar{\mu}^{-3/2} \eta^2 \alpha^{-4} \nu^{-1} \unit{erg} \unit{s^{-1}} \unit{cm^{-3}}$
in volume-averaged power per unit volume due to free-free emission from gas
with $T > 0.63 \bar{\mu} \nu^2 \unit{keV}$ for large-box $C2.5$ (solid), averaged over one
orbit, and large-box $S$ (dotted).
The peak power per volume is higher when the stars are orbiting than when they are stationary.
\label{fig:highd-emiss}}
\end{figure}

\begin{figure}
\epsscale{1.0}
\plotone{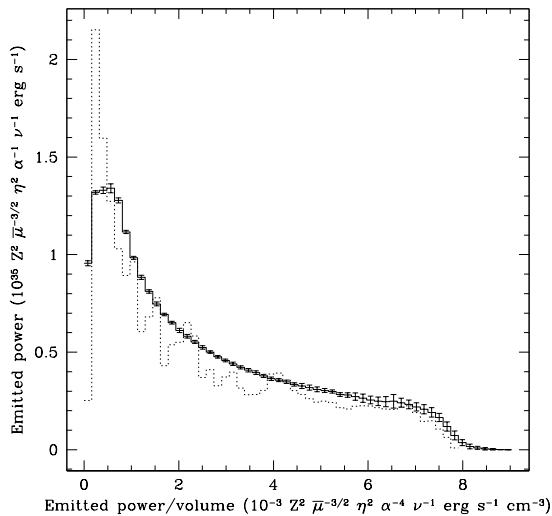}
\figcaption{
Histogram of power emitted from bins of width
$\Delta \Lambda_{ff} = 1.25 \times 10^{-4} Z^2 \bar{\mu}^{-3/2} \eta^2 \alpha^{-4} \nu^{-1} \unit{erg} \unit{s^{-1}} \unit{cm^{-3}}$
in volume-averaged power per unit volume due to free-free emission from
gas with $T > 0.63 \bar{\mu} \nu^2 \unit{keV}$ for large-box $C2.5$ (solid), averaged
over one orbit, and large-box $S$ (dotted).
There is more integrated emission from low power per volume but less from high power per volume when the stars are stationary compared to when they are orbiting.
\label{fig:highd-emtot}}
\end{figure}

Figure \ref{fig:highd-emiss} shows volume as a function of power per unit
volume due to free-free emission from gas with $T > 0.63 \bar{\mu} \nu^2
\unit{keV}$.  In $C2.5$, gas with
$\Lambda_{ff} > 4.5 \times 10^{-3} Z^2 \bar{\mu}^{-3/2} \eta^2 \alpha^{-4} \nu^{-1} \unit{erg} \unit{s^{-1}} \unit{cm^{-3}}$,
all has $T > 0.875 \bar{\mu} \nu^2 \unit{keV}$, therefore using that cutoff
instead of $T > 0.63 \bar{\mu} \nu^2 \unit{keV}$ will not cause us to miss the
high-emissivity gas.  While most of the post-shock volume at low temperatures
emits little power per unit volume, the integrated power from this volume, as
shown by Figure \ref{fig:highd-emtot}, can still contribute significantly to
the total emission.  The power emitted from gas with $T > 0.875 \bar{\mu} \nu^2
\unit{keV}$ is only $79\%$ of that emitted from all gas with $T > 0.63
\bar{\mu} \nu^2 \unit{keV}$ in the large-box $C2.5$.  Similarly, the power
emitted from gas with $T > 0.875 \bar{\mu} \nu^2 \unit{keV}$ is only $82\%$ of
that emitted from all gas with $T > 0.63 \bar{\mu} \nu^2 \unit{keV}$ in the
large-box $S$.  The quantitative change between observables measured will
depend on the cutoff used, but since we are using the same cutoff for each
simulation when making direct comparisons, we will still get a qualitative feel
for the effect as well as an esimate of its magnitude.

\begin{figure}
\epsscale{1.0}
\plotone{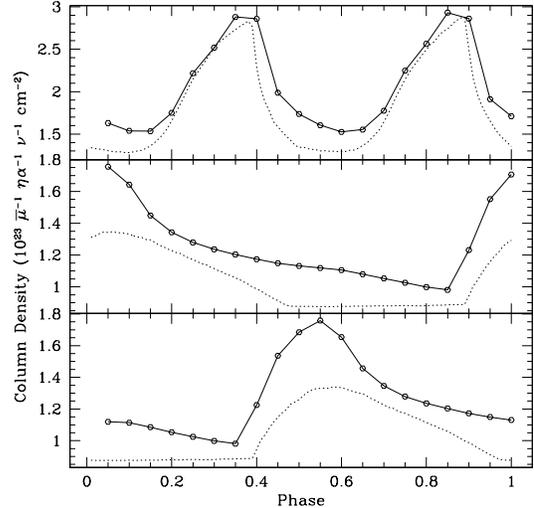}
\figcaption{
Column density on the line of sight described in \S\ref{sec:diagnostics} to the center of mass (top), primary star (middle), and secondary star (bottom), for $C2.5$ in a large box (solid) and small box (dotted).
\label{fig:highd-column}}
\end{figure}

Column density, as described in \S\ref{sec:diagnostics}, is plotted in
Figure \ref{fig:highd-column}.  The column density to the center of mass
is determined mostly by the density of the post-shock material and the
thickness of the post-shock region along the line of sight.  At phases
0.5 and 1.0, the line of sight falls in the cut plane shown in Figure
\ref{fig:den640}e, at an angle 25$^\circ$ from vertical.  At phases 0.25
and 0.75, the line of sight falls in the cut plane shown in Figure
\ref{fig:den640}f.  The column density at phases
0.25 and 0.75 is higher than at 0.5 and 1.0 because, although the line
of sight passes through slightly less post-shock material, it passes
through more post-shock material close to the center of mass of the
system, where the mass density is highest.  When a smaller box is used,
the column density to the center of mass changes most near the minima.
The post-shock gas that was near the edge of the box in the large-box
runs is not visible in the small-box runs.  Even though the values change,
the qualitative features remain.

The column density to the primary and secondary when the orbit is
circular and the winds are identical is essentially the same but with a
phase shift of 0.5.  It is highest near when the stars are on the far
side of the center of mass relative to the observer, near phase 0.0
for the primary and 0.5 for the secondary.  When the stars are equally
distant from the observer, the column density is higher to the star
that is moving toward the observer, the primary at phase 0.25 and the
secondary at phase 0.75, because the post-shock material leads the star.
When a smaller box is used, the column density to the stars is affected
at all phases.  Instead of continuously decreasing after one peak until
the next peak is approached, in a small box the column density flattens
out well before the next peak.

\subsection{Circular-orbit Parameter Survey}\label{sec:circsurvey}

\begin{figure}
\epsscale{1.0}
\plotone{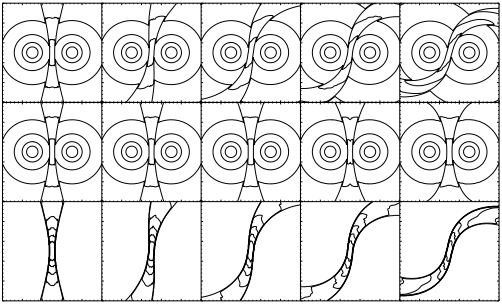}
\figcaption{
Logarithmic density contours in the orbital plane (top) and normal to the orbital plane (middle) and linear temperature contours in the orbital plane (bottom) for (left to right) small-box circular-orbit simulations $S$, $C10$, $C5$, $C3.5$, and $C2.5$.
The smaller size of the computational domain ($L = 2.5 \alpha \unit{AU}$) means
the spiral pattern seen at larger distances from the stars is not captured.
\label{fig:contours}}
\end{figure}

Figure \ref{fig:contours} shows density contours both in the orbital
plane and normal to it, as well as temperature contours in the orbital
plane, for five different simulations that survey the ratio of
wind to orbital velocity.  All use stars on circular orbits in a box
of size $L = 2.5 \alpha \unit{AU}$.  The stars are
moving counter-clockwise, as before.  The shock fronts curve around the
stars more and more as the orbital period decreases, both in the orbital
plane and normal to it.  The smaller box crops off the spiral patterns
visible in the large-box simulations (Figures \ref{fig:den640} and
\ref{fig:temper640}).  The maximum post-shock density shows a nearly
monotonic decrease of $\sim 6\%$ as the velocity ratio decreases to
$V_w/V_o = 2.5$, while the maximum post-shock temperature doesn't change
significantly.

The power radiated by free-free emission from gas with $T > 0.875 \bar{\mu}
\nu^2 \unit{keV}$ increases as orbital period decreases.  The time-average of
the power emitted from $C10$ is only $0.085\%$ higher than from $S$.
This justifies using $C10$ as a reference model instead of $S$ in the
discussion to follow since $S$ is stationary on the grid and will have
more grid noise.  The time-average of the power emitted from $C5$ is
$1.9\%$ higher, from $C3.5$ is $4.0\%$ higher, and from $C2.5$ is $7.7\%$
higher than from $C10$.

\begin{figure}
\epsscale{1.0}
\plotone{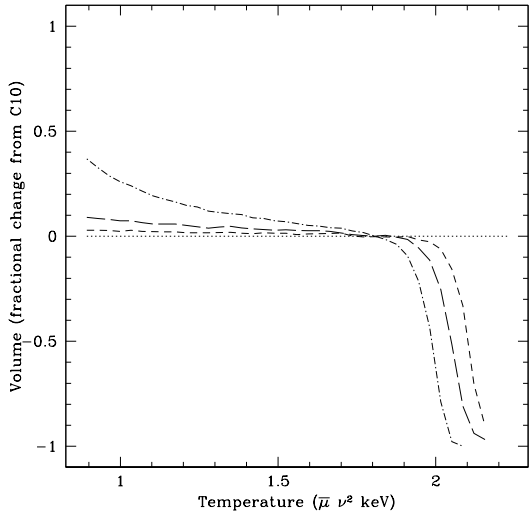}
\figcaption{
Histogram of fractional difference in volume, averaged over one orbit, within post-shock temperature bins of width $\Delta T = 0.0212 \bar{\mu} \nu^2 \unit{keV}$ for simulations $C5$ (short dash), $C3.5$ (long dash), and $C2.5$ (dash dot) relative to $C10$ (dotted).
There is more volume at lower temperatures and less at higher temperatures when the orbital period is shorter.
\label{fig:avgtemp}}
\end{figure}

Figure \ref{fig:avgtemp} shows the fractional difference in post-shock
volume, averaged over one orbital period, as a function of post-shock
temperature for the circular-orbit simulations when compared to simulation
$C10$, with the lowest non-zero orbital velocity.  For temperatures below $T
\approx 1.75 \bar{\mu} \nu^2 \unit{keV}$, the volume is larger for simulations
with shorter orbital periods because the post-shock temperatures are higher
on the leading side of the post-shock region than they would be if the
stars were stationary.  For higher temperatures, the volume is smaller
for simulations with shorter orbital periods because the wind impacts
the shock front along the line of centers at a more oblique angle.

We histogram the volume in bins of width
$\Delta \Lambda_{ff} = 1.25 \times 10^{-4} Z^2 \bar{\mu}^{-3/2} \eta^2 \alpha^{-4} \nu^{-1} \unit{erg} \unit{s^{-1}} \unit{cm^{-3}}$
in power per unit
volume due to free-free emission from gas with $T > 0.875 \bar{\mu} \nu^2
\unit{keV}$ for $C5$, $C3.5$, and $C2.5$ compared to $C10$.  $C10$ has the
highest maximum power per unit volume, although $C3.5$ and $C2.5$ have more
volume near that value.  At intermediate values of power/volume, in the range
$1.5 \times 10^{-3} < \Lambda_{ff,cgs} Z^{-2} \bar{\mu}^{3/2} \eta^{-2} \alpha^4 \nu < 4.5 \times 10^{-3}$,
the volume in
a given bin is an average of $2.7\%$ higher for $C5$, $6.8\%$ higher for
$C3.5$, and $15\%$ higher for $C2.5$ than for $C10$.  The power emitted
from the trailing side of the post-shock region is higher than from the
leading side when there is orbital motion.

\begin{figure}
\epsscale{1.0}
\plotone{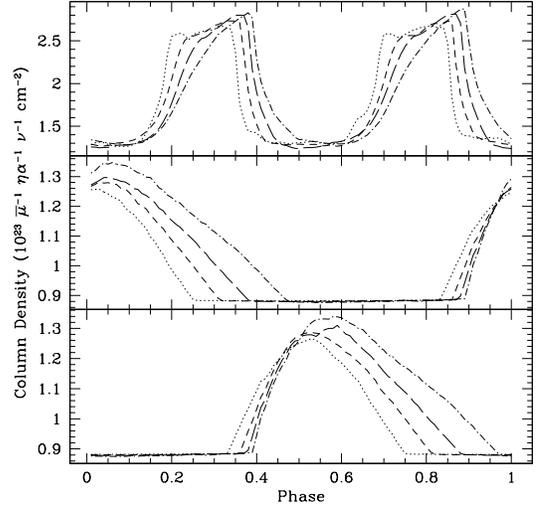}
\figcaption{
Column density on the line of sight described in \S\ref{sec:diagnostics} to the center of mass (top), primary star (middle), and secondary star (bottom).
\label{fig:column}}
\end{figure}

The ratio of wind velocity to orbital velocity impacts the column density to the hot post-shock gas as well.
Figure \ref{fig:column} shows column density for each of the circular-orbit simulations.
The column density to the center of mass of the system, where there will be the hardest X-ray emission from the post-shock gas, peaks twice per orbital period.
While the column density should peak when the stars are half way between conjunctions when there is no curvature of the shocked gas due to the Coriolis force, as the orbital period decreases and the Coriolis force becomes stronger, the peak column densities are delayed and the peaks become less symmetrical, increasing more slowly and then decreasing more rapidly.
The peak is delayed by a phase of 0.07 for $C2.5$ compared to $C10$ and the difference between the peak and minimum column density increases by $13\%$ between $C10$ and $C2.5$.
The FWHM remains roughly constant as the orbital velocity is varied.

When the orbital period is very long, the column density to the stars is nearly symmetrical in time, but as the orbital period decreases and the curvature of the shocked region becomes stronger, the column density becomes very asymmetrical.
As the orbital period decreases, the column density increases more rapidly leading up to the peak and then decreases less rapidly after the peak, causing an increase in the FWHM of the peaks of $31\%$ from $C10$ to $C2.5$.
The column density of the star that is father away from the observer increases quickly as the stars approach conjunction.
The peak column density should occur at conjunction when the shock fronts aren't curved by the Coriolis force, but when they are curved, the peak occurs after conjunction.
For $C10$, the peak occurs at a phase only 0.02 past conjunction, but for $C2.5$, the peak is delayed to a phase 0.07 past conjunction.
The difference between the peak and minimum column density increases by $22\%$ between $C10$ and $C2.5$.
The fraction of the orbit during which the column density is significantly above its minimum value is higher for systems with shorter orbital periods.
It is only $41\%$ for $C10$ but increases to $57\%$ for $C2.5$.

\subsection{Unequal Winds on Elliptical Orbits}

To show how a combination of an elliptical orbit and unequal winds
affects the shock structure, we present images of a large-box ($L = 5 \alpha
\unit{AU}$) elliptical-orbit simulation with unequal mass-loss rates.  The
mass-loss rate of the primary has been increased by a factor of 1.5 compared
to the secondary and the eccentricity of the orbit is $e = 0.2$.

\begin{figure}
\epsscale{0.3}
\plotone{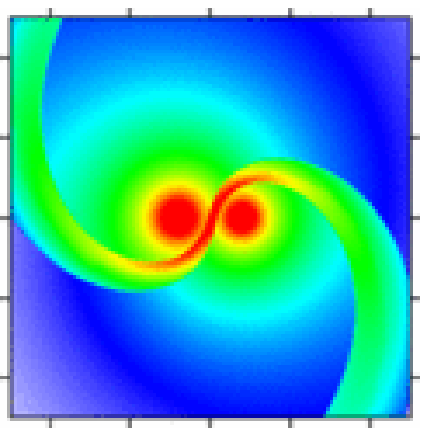}
\plotone{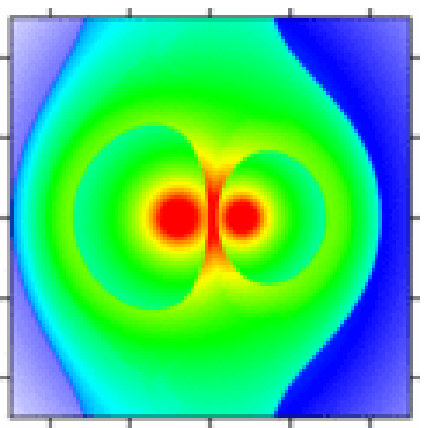}
\plotone{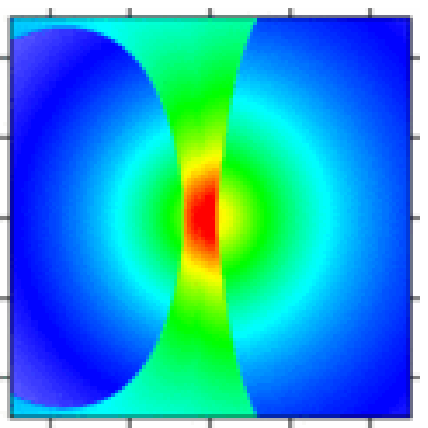} \\
\plotone{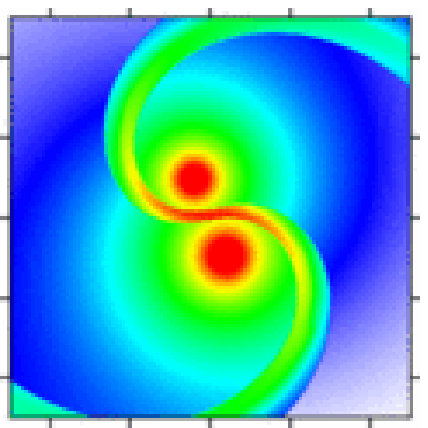}
\plotone{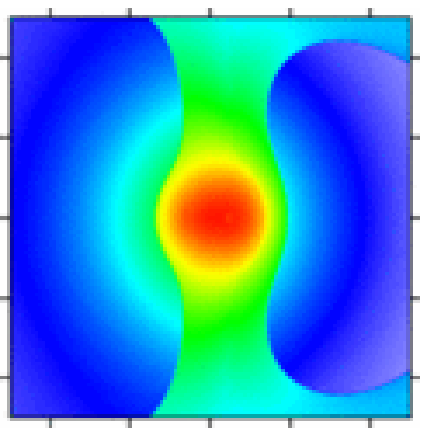}
\plotone{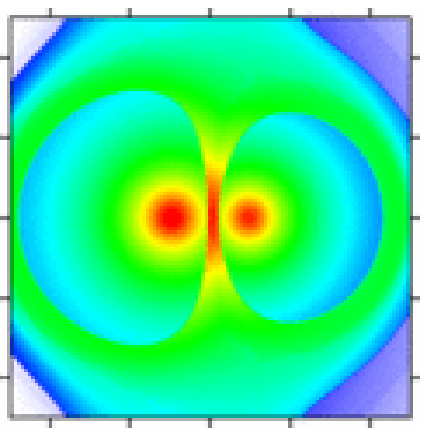} \\
\plotone{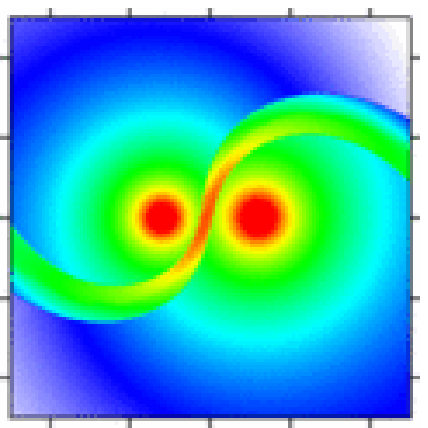}
\plotone{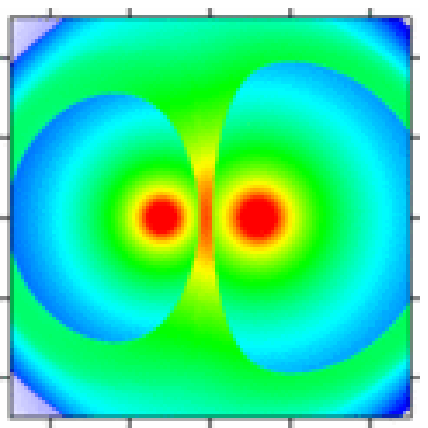}
\plotone{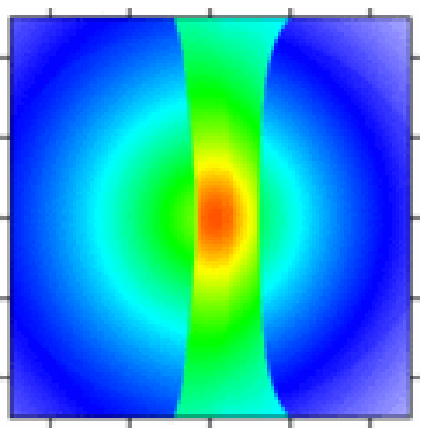}
\figcaption{
Density in large-box simulation with unequal mass-loss rates on elliptical orbits
(\emph{a}) in the orbital plane, 
(\emph{b}) along the semi-major axis normal to the orbital plane, and
(\emph{c}) normal to the previous two slices, at phase 0.5.
The same slices, 
(\emph{d}), (\emph{e}), and (\emph{f}), respectively, at phase 0.75, and
(\emph{g}), (\emph{h}), and (\emph{i}), respectively, at phase 1.0.
The color scale is logarithmic, with $\alpha^2 \nu \eta_{-1} \rho_{-16}$ from 1 (white) to 700 (red).
The tick marks are spaced $\alpha \unit{AU}$.
\label{fig:den640ecc}}
\end{figure}

\begin{figure}
\epsscale{0.3}
\plotone{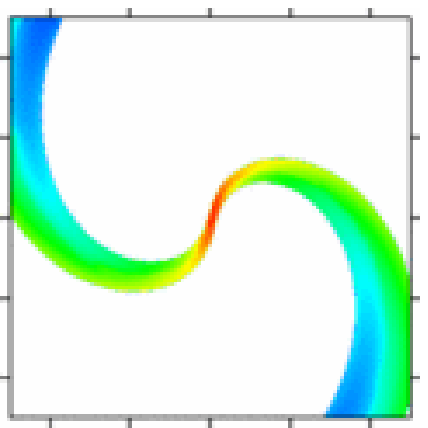}
\plotone{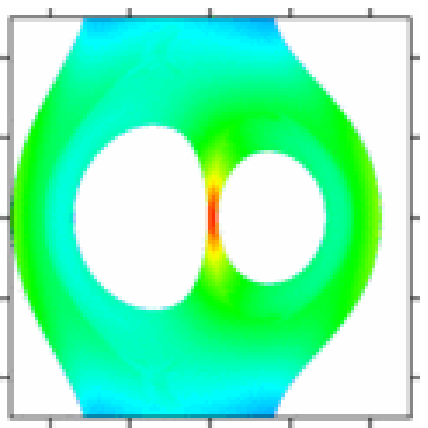}
\plotone{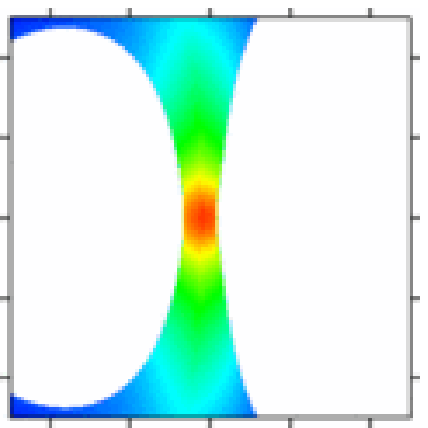} \\
\plotone{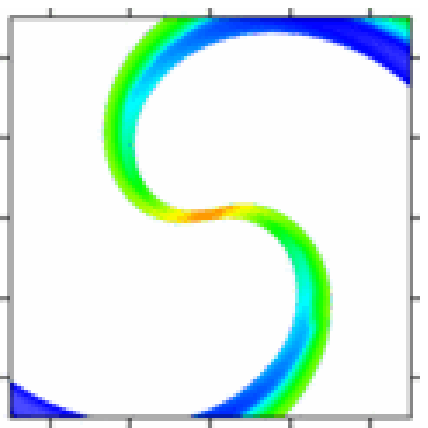}
\plotone{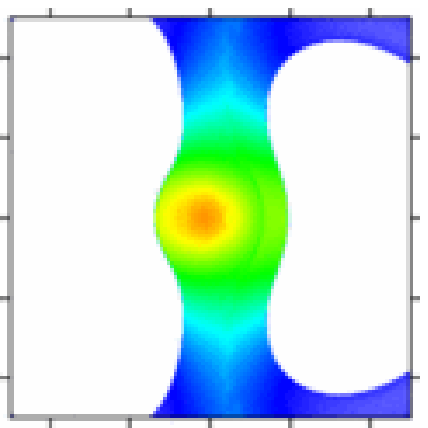}
\plotone{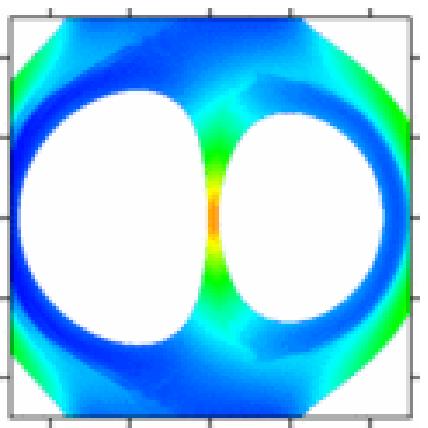} \\
\plotone{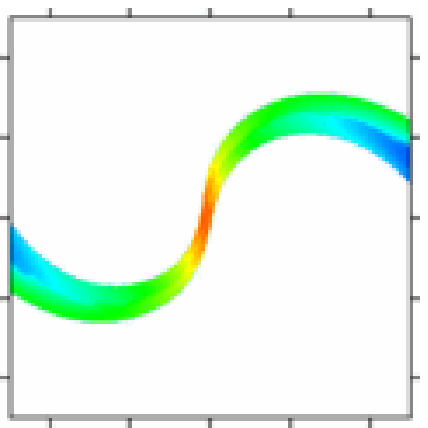}
\plotone{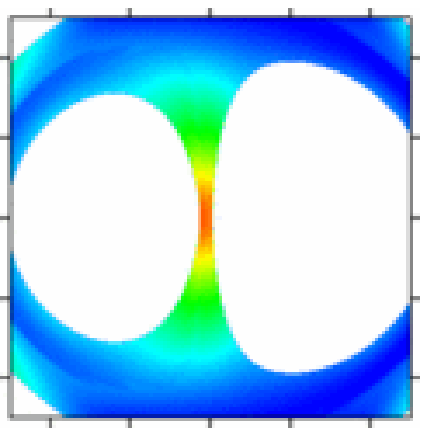}
\plotone{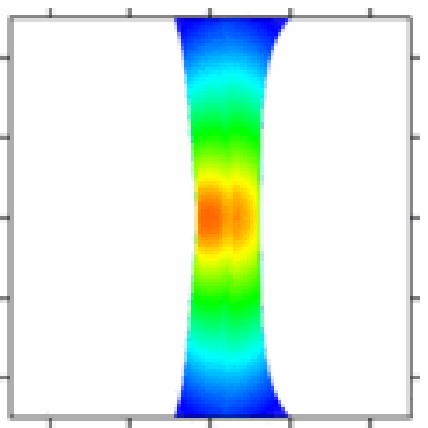}
\figcaption{
Temperature in large-box simulation with unequal mass-loss rates on elliptical orbits
(\emph{a}) in the orbital plane, 
(\emph{b}) along the semi-major axis normal to the orbital plane, and
(\emph{c}) normal to the previous two slices, at phase 0.5.
The same slices, 
(\emph{d}), (\emph{e}), and (\emph{f}), respectively, at phase 0.75, and
(\emph{g}), (\emph{h}), and (\emph{i}), respectively, at phase 1.0.
The color scale is logarithmic, with $\bar{\mu}^{-1} \nu^{-2} T_{\unit{keV}}$ from 0.05 (white) to 2.5 (red).
The tick marks are spaced $\alpha \unit{AU}$.
\label{fig:temper640ecc}}
\end{figure}

Figure \ref{fig:den640ecc} shows the density in three orthogonal cut planes
that pass through the center of mass of the system.  These planes are shown
at three different phases of the orbit.  Figure \ref{fig:temper640ecc}
shows the temperature in the same three planes and phases.  The curvature
of the shocks varies as a function of phase due to the elliptical orbit
and is also different for the two stars.  The curvature is stronger around
the secondary star, with the lower mass-loss rate wind.  The post-shock
material between the stars is also shifted toward the secondary.

We will now investigate separately the effect on post-shock temperatures, free-free emission, and column density of elliptical orbits and unequal winds.
In \S\ref{sec:ellipresults}, we present the analysis of a simulation with identical winds on an elliptical orbit.
In \S\ref{sec:uneqresults}, we analyze two simulations, one with differing mass-loss rates and the other with differing stellar wind velocities, both on circular orbits.
These simulations are run in a smaller box, with $L = 2.5 \alpha \unit{AU}$.

\subsection{Identical Winds on Elliptical Orbits}\label{sec:ellipresults}

We consider here stars with identical winds on elliptical orbits, as described in \S\ref{sec:systems}.
Since we do not model the radiative driving of the winds, we find that the highest post-shock temperatures are caused by the wind that leaves the masks when the star have their greatest velocity towards each other.
Pittard (1998), however, found that the highest post-shock temperatures occur near apastron when the winds have had the longest distance to accelerate.
Using equation (\ref{eq:beta}), we can compare the magnitude of these two effects.
Assuming the physical parameters for the binary O-star system described in
\S\ref{sec:systems} and taking the stellar radius to be $R_* = 20 R_\odot$, we
find that the wind velocity at the stagnation point when the stars are at
apastron should be $6.8\%$ higher than when the stars are at periastron.  We
also find that the combined radial and wind velocity is $18\%$ higher when
the stars are moving toward each other than when they are moving away
from each other with their highest relative velocity.  The magnitudes
of these effects are comparable.

\begin{figure}
\epsscale{1.0}
\plotone{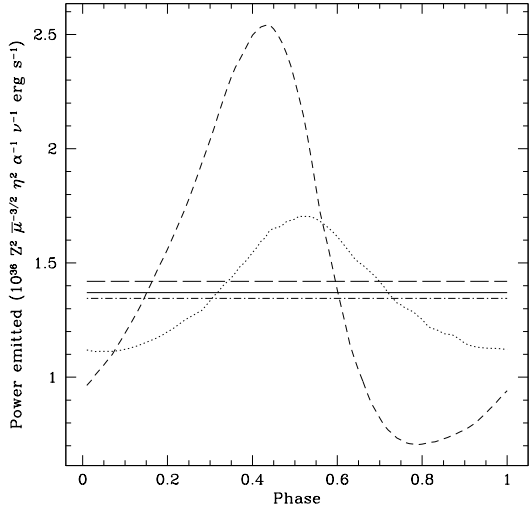}
\figcaption{
Frequency-integrated free-free power emitted from gas with $T > 1.155 \bar{\mu} \nu^2 \unit{keV}$ as a function of phase from $SE2.5$ (dotted) and $E2.5$ (short dash).
The time-average for $C2.5$ (solid), $SE2.5$ (dash dot), and $E2.5$ (long dash) are also shown.
Varying the stellar separation without full orbital motion is not a good approximation to an elliptical orbit.
\label{fig:emsumecc}}
\end{figure}

We use a temperature cut of $T > 1.155 \bar{\mu} \nu^2 \unit{keV}$ for the
elliptical-orbit analysis.  The maximum in power radiated by free-free emission
from gas with $T > 1.155 \bar{\mu} \nu^2 \unit{keV}$ occurs before periastron,
at phase 0.44, when full orbital motion is allowed, as shown in Figure
\ref{fig:emsumecc}.  While the mean power emitted differs by only $3.6\%$
between the circular and elliptical cases, the instantaneous value varies
greatly.  The maximum is $79\%$ higher and the minimum is $50\%$ lower than the
mean.  When we instead vary the stellar separation without full orbital motion
($SE2.5$), we find that the maximum in power emitted occurs after periastron.
The time delay from periastron is due to the wind released by the star
taking time to reach the shock front.  The mean for $SE2.5$ is $5.3\%$
lower than for $E2.5$.  The maximum for $SE2.5$ is only $27\%$ higher than
the mean and the minimum is only $17\%$ lower.  The curve for $SE2.5$
is symmetrical in time while that for $E2.5$ is not.  We will discuss
the reason for this below.

While the stellar separation is the same at phase $\phi$ as at phase
$1.0 - \phi$, the stellar radial velocities differ in sign, so we do
not expect the post-shock temperature distribution to be symmetrical
in time.  We instead find that, while the velocity of the stars towards
each other is increasing, the high-temperature tail of the distribution
drops off more steeply than for a circular orbit and, conversely, when
the stellar radial velocity is increasing, the high-temperature tail
drops off less steeply.  The sign of the change in volume at a given
temperature switches first for low temperatures and later for higher
temperatures.

\begin{figure}
\epsscale{1.0}
\plotone{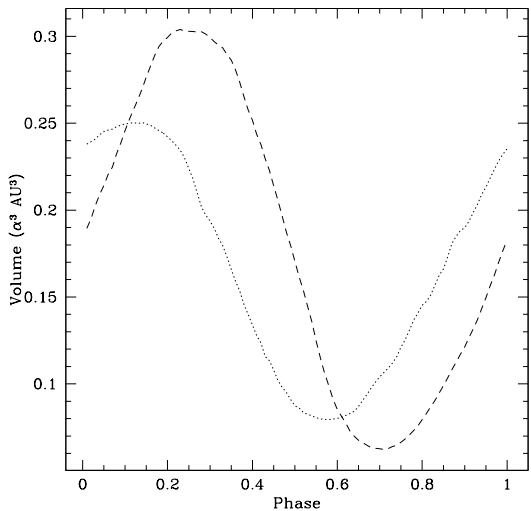}
\figcaption{
Volume with $T > 1.155 \bar{\mu} \nu^2 \unit{keV}$ as a function of phase for $SE2.5$ (dotted) and $E2.5$ (short dash).
\label{fig:volecc}}
\end{figure}

While the total volume with $T > 1.155 \bar{\mu} \nu^2 \unit{keV}$, shown in Figure
\ref{fig:volecc}, in $SE2.5$ is correlated with the separation between the
stars, with more volume for larger separations, in $E2.5$ it is correlated
with the stellar radial velocity, with the most volume when the stars are
moving towards each other with the highest velocities.  At apastron the
shock thickness along the line of centers is approximately $14\%$ larger
than for the circular-orbit case and at periastron it is approximately
$14\%$ smaller.  The phase delay for the gas leaving the mask to reach the
shock in $SE2.5$ is typically between 0.02 and 0.04.

Although the maximum and minimum post-shock temperatures do not occur near
periastron and apastron, respectively, in our simulation, the maximum and
minimum peak power/volume due to free-free emission do.  The maximum peak
power/volume in $E2.5$ matches that in $C2.5$ near phases 0.27 and 0.71.  At
phase 0.27, the power/volume distribution is similar for $C2.5$ and $E2.5$
except at low values, where there is more volume from $T > 1.155 \bar{\mu}
 \nu^2 \unit{keV}$ gas for $E2.5$.  At phase 0.71, the power/volume
distribution is again similar except at low values, where there is now less
volume from $T > 1.155 \bar{\mu} \nu^2 \unit{keV}$ gas for $E2.5$.

\begin{figure}
\epsscale{1.0}
\plotone{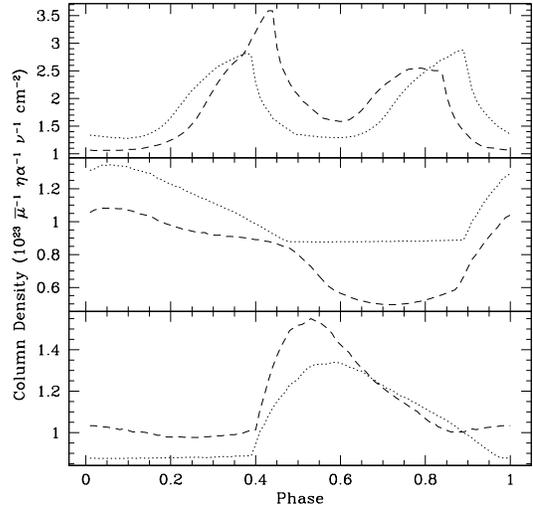}
\figcaption{
Column density on the line of sight described in \S\ref{sec:diagnostics} to the center of mass (top), primary star (middle), and secondary star (bottom) for $C2.5$ (dotted) and $E2.5$ (short dash).
\label{fig:colecc}}
\end{figure}

In Figure \ref{fig:colecc}, we compare the column density as a function
of phase for $C2.5$ and $E2.5$.  The projection of the line of sight onto
the orbital plane runs parallel to the semi-major axis of the elliptical
orbit.  When the orbit is circular, the two peaks in column density
when looking toward the center of mass of the system are identical,
but when the orbit is elliptical, the peak before periastron is higher
and the peak after is lower.  The minimum before periastron is lower
and the minimum after is higher.  The difference between the peak and
minimum column density is $60\%$ higher for the elliptical orbit than
for the circular orbit.

Since the separation between the stars is no longer the same half an
orbit apart, the column density to the primary and secondary no longer
share the same phase-shifted time dependence even though the individual
winds are still identical.  The column density to the primary is now
lower at all phases than it was in the circular-orbit case, while the
column density to the secondary is higher for most, but not all, phases.
Where the column density to the primary is nearly flat under a circular
orbit, it now varies with time due to the changing stellar separation.
The secondary has a narrower peak near periastron and isn't quite
as flat after apastron.  The difference between the minimum and peak
column densities is $23\%$ higher for the elliptical orbit than for the
circular orbit.

\subsection{Unequal Winds on Circular Orbits}\label{sec:uneqresults}

When we consider stars with differing wind parameters on circular orbits,
we find that the thickness of the post-shock region along the line
of centers does not change.  The post-shock region is shifted toward
the secondary, though, giving a distance ratio from the star centers
to the center of the post-shock region of 1.2 for $CM2.5$ and 1.3 for
$CW2.5$.  We expect that modifying the wind velocity will change the
temperature distribution, but that modifying the mass-loss rate will not.
This follows from the equation for energy per unit mass,
\begin{equation}
\frac{E}{\rho} = \frac{1}{\gamma - 1} \frac{P}{\rho} + \frac{1}{2} v^2.
\end{equation}
Since $T \propto P / \rho$, this implies that $T \propto v^2$.  The wind
density is set by both the mass-loss rate and wind velocity, however,
so we expect that modifying either one will cause a change in the
emissivities.

We histogram the volume in bins of width $\Delta T = 0.0212 \bar{\mu} \nu^2
\unit{keV}$ in post-shock temperature using a temperature cut of $T > 0.9625
\bar{\mu} \nu^2 \unit{keV}$ for the unequal wind analysis.  As expected, the
temperature distribution changes very little when only the mass-loss rate is
modified.  The total post-shock volume above $T = 0.9625 \bar{\mu} \nu^2
\unit{keV}$ in $CM2.5$ is lower by $1.3\%$ compared to $C2.5$.  The difference
in volume in each bin is roughly constant over $0.9625 < T_{\unit{keV}}
\bar{\mu}^{-1} \nu^{-2} < 1.9425$.  When the wind velocity is modified,
however, there is much
less volume at a given temperature above the cutoff, and the peak temperature
is slightly lower.  Since the secondary star has a lower wind velocity, the
temperature of the secondary's shocked wind is much lower than that of the
primary.  Due to the relation above, we expect the shocked wind of the
secondary to fall below the temperature cutoff.  $CW2.5$ should then have the
same temperature distribution above $T = 0.9625 \bar{\mu} \nu^2 \unit{keV}$ as
$C2.5$, but with the volume in all bins reduced by a constant factor.  Since
there is $55\%$ less post-shock volume at all temperatures above $T = 0.9625
\bar{\mu} \nu^2 \unit{keV}$, there is also $55\%$ less volume in each bin in
the range $0.9625 < T_{\unit{keV}} \bar{\mu}^{-1} \nu^{-2} < 1.9425$.

We histogram the power per unit volume from gas with $T > 0.9625 \bar{\mu}
\nu^2 \unit{keV}$ and find that the volume peaks at low
power/volume.  $CM2.5$ has less volume per bin than $C2.5$ except at the very
lowest and highest bins.  With the exception of near and above the highest
power/volume found in $C2.5$, $CW2.5$ also has less volume per bin.  As with
the temperature distribution, the reason for this is that the secondary's
shocked wind is at too low a temperature to be included in the analysis,
resulting in less emission.  $CW2.5$ has a much higher maximum power/volume
than $C2.5$ and $CM2.5$.  The time-averaged power radiated by free-free
emission from gas with $T > 0.9625 \bar{\mu} \nu^2 \unit{keV}$ is $0.33\%$
lower for $CM2.5$ and $64\%$ lower for $CW2.5$ than for $C2.5$.  The standard
deviation in this value is $0.58\%$ of the mean for $C2.5$, $0.30\%$ for
$CM2.5$, and $2.8\%$ for $CW2.5$.  The KH instability appears to be strongest
in the case where the wind velocities are unequal.

\begin{figure}
\epsscale{1.0}
\plotone{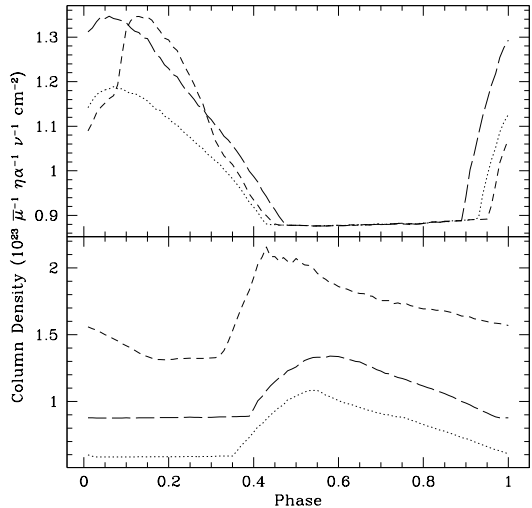}
\figcaption{
Column density on the line of sight described in \S\ref{sec:diagnostics} to the primary star (top) and secondary star (bottom) for $CM2.5$ (dotted), $CW2.5$ (short dash), and $C2.5$ (long dash).
\label{fig:coluneq}}
\end{figure}

The column density should be different for both the modified wind
velocity and mass-loss rate due to changes in the density.  The column
density over a full orbital period for these two cases is shown in
Figure \ref{fig:coluneq}.  The column density to the center of mass is not
shown here because the post-shock region is no longer centered on that
location.  The mass density of the secondary is lower in
$CM2.5$ and higher in $CW2.5$ compared to $C2.5$, resulting in a lower
and higher column density, respectively, to the secondary at all phases.
When the column density to the primary is at its minimum value, it is
the same for the equal and unequal wind simulations because the line of
sight passes only through the wind of the primary, which is unchanged.
When the line of sight passes through both winds, the column density is
higher in $CW2.5$ and lower in $CM2.5$.  The column density is at its
minimum value for the shortest amount of time when the winds are equal.
Since the shock will wrap more tightly around the secondary than the
primary, the line of sight will pass through only the wind of the primary
for a larger range of phases when the winds are unequal.

\section{DISCUSSION}\label{sec:conclusion}

WR 20a is a binary system of nearly-identical Wolf-Rayet stars, likely
type WN6ha, on circular orbits with a period of 3.686 days (Rauw et
al. 2005).  The primary has an estimated mass of 82.7 M$_\odot$ and the
secondary 81.9 M$_\odot$, giving a mass ratio of 1.01.  This gives a
semi-major axis for the system of $a = 0.256 \unit{AU}$, or $\alpha = 0.256$.
If we assume instead a mass ratio of 1.0, taking $M_1 = M_2 = 82.3 M_\odot$,
and use Rauw's assumed wind velocity at the stagnation point of $500 \unit{km}
\unit{s^{-1}}$ for each star, or $\nu = 0.5$, this would give $V_w/V_o \approx
1.3$.  We also take Rauw et al.'s smooth-wind mass-loss rate of $\dot{M}_1
= \dot{M}_2 = 2.5 \times 10^{-5} M_\odot \unit{yr^{-1}}$, or $\eta = 2.5$.
We also assume $\bar{\mu} = 1.4$ (WN).

Although $V_w/V_o$ for this system is smaller than any ratio we modeled, we
can still put a lower bound on the importance of the Coriolis force by
comparing models $S$ and $C2.5$ for this system.  The wind mach numbers we
used in our simulations are smaller than the actual values for this system,
but since we are already in the strong-shock limit, this shouldn't have a
qualitative effect on our results.  For $C2.5$, we find the power radiated
by free-free emission to be $5.5 \times 10^{37} Z^2 \unit{erg} \unit{s^{-1}}$
for $T > 0.22 \unit{keV}$.  For $S$, we find only $4.8 \times 10^{37} Z^2
\unit{erg} \unit{s^{-1}}$.  This is a difference of $13\%$, even though we
used a larger $V_w/V_o$ than is appropriate.  Without taking wind
acceleration or radiative effects into account, we find that the orbital
motion has a significant, although not easily measured, effect on the power
radiated by free-free emission.  Unfortunately, by eq. (\ref{eq:alpharad}),
we find $\alpha_{rad} = 373 \gg \alpha$, meaning this system is actually
quite radiative.

Had our simulations incorporated radiative driving of the wind, resulting in
positive net wind acceleration, the increased wind velocity would likely 
cause the thickness of the post-shock region to decrease, with the largest 
effect furthest from the line of centers.  The modified velocity profile would 
also give an effective $V_w/V_o$ that increases with distance from the line of 
centers, which might also cause the shocks to wrap more tightly around the 
stars.  The former effect would cause the wind to impact the shock front at a 
more oblique angle, while the latter would have the same effect on the 
trailing side of the shock and the opposite effect on the leading side.  These 
two effects, combined with the increased wind velocity, makes determining the 
change in the post-shock temperature distribution difficult.

For an isolated star, gravity acts against the radiative driving of the wind, 
leading to a different velocity profile in the unshocked wind, with lower 
velocities than those expected without gravity.  When these modified winds 
collide, the thickness of the post-shock region far from the line of centers 
would be increased significantly, leading to a larger post-shock volume.  
Gravity acts not only on the unshocked winds, but also on the post-shock gas, 
however.  In the post-shock region between a pair of equal-mass stars that are 
held stationary, the force due to gravity would pull the post-shock gas toward 
the stagnation point.  This would lead to an enhanced density near the plane 
normal to the line of centers, which would likely result in a decrease in 
temperature near the contact surface.  This discussion is based on simple
expectations; a complete exploration of the effect of gravity on the post-shock
flow will require 3D simulations in which it is included.

Since a higher ratio of $V_w/V_o$ in our simulations corresponds to an 
increased stellar mass, the effects due to gravity will increase with 
decreasing orbital period.  We find that the escape velocity for 
$V_w/V_o = 2.5$ at a distance of $1.25 \alpha \unit{AU}$ from the stagnation 
point is roughly equal to the unshocked wind velocity at the stagnation point.
Clearly gravity will have a strong influence on the post-shock region in the
absence of radiative driving for systems with a large Coriolis force, therefore 
conducting simulations that include both the effects of gravity and radiative 
driving is the logical next step.  This, however, is beyond the scope of this 
paper.

Our analysis in \S\ref{sec:circsurvey} shows that, although the orbital
motion has little effect on the peak temperature and density along the
line of centers between the stars, it can have a significant effect on the
temperature and emissivity of the gas elsewhere in the post-shock region.
It also impacts the light curves by changing the time-dependence of the
attenuation and allows for the K-H instability even when the winds are
identical.  The shape of the shocked region changes substantially when
orbital motion is added, causing the distance between the stars and the
shock fronts to decrease.  The shocks also have a rotational velocity once
orbital motion is considered, causing a difference in the shock strength
on the leading and trailing sides.  While we did not consider radiative
effects or gravitational forces that would cause the wind velocity to vary
with distance from the star, adding these effects will likely have a large
impact on the system.  It therefore would be fruitful in the future to 
consider fully three-dimensional hydrodynamical models which include both 
radiative driving and gravity.

\acknowledgements
This work was supported by the DoE through grant DE-FG52-06NA26217.
Simulations were performed in the IBM Blue Gene at Princeton University, and
on computational facilities supported by NSF grant AST-0216105.

\appendix

\section{Wind Generation Method}\label{sec:windgen}

The wind is generated by imposing the wind solution, described in
\S\ref{sec:windsol}, onto a sphere of radius $r_0$, called the masked
region, before every time step.  The values of the density, pressure,
and velocity vector within each cell in the masked region are computed using
the volume-average of the analytic wind solution over the cell, for example
for the density
\begin{equation}\label{eq:rhoavg}
\langle\rho\rangle = \frac{\int \rho(r) dV}{\int dV},
\end{equation}
where $dV=dx \unit{} dy \unit{} dz$ is the volume of the cell and $r^{2} = x^{2}+y^{2}+z^{2}$.
A similar formula is used for the pressure.
The radial unit vector, ${\bf r}$, is also averaged in this manner,
and is used as the direction of the wind velocity in each grid cell within the
mask in the reference frame of the star.  Since the stars are moving,
in the reference frame of the grid the wind
velocity is then
\begin{equation}
{\bf v} = v_w \langle{\bf r}\rangle + {\bf v}_*,
\end{equation}
where ${\bf v}_*$ is the stellar velocity.
The momentum density vector and the energy density are then calculated from the cell-averaged pressure, mass density, and velocity using
\begin{equation}
{\bf p} = \langle\rho\rangle {\bf v}
\end{equation}
and
\begin{equation}\label{eq:energyfn}
E = \frac{\langle{P}\rangle}{\gamma - 1} + \frac{1}{2} \langle\rho\rangle {\bf v}^2.
\end{equation}
The singularity in the density and pressure at the center of the star and the issue of dynamic range in the simulation are handled simultaneously by imposing a maximum density and pressure, $\rho_{max} = 4 \rho_0$ and $P_{max} = (2)^{10/3} P_0$, before averaging.
These maxima fall at $r = 0.5 r_0$.

The integrals of the form of eq. (\ref{eq:rhoavg}) are approximated
using quadratures by dividing each cell into $10^3$ segments and then
using the midpoint rule.  Since space is discretized on the grid, cells
which fall only partially within the masked region are treated as such.
The new grid values for these cells are a linear combination of the
previous value and the masked value,
\begin{equation}
\vec{q}_{n+1} = f \vec{q}_{mask} + (1-f) \vec{q}_n,
\end{equation}
where $\vec{q} = (\rho,E,p_1,p_2,p_3)$ and $f$ is the fraction of the cell's volume covered by the mask.
The masks are moved over the grid in the appropriate orbit using an N-body integrator.


\begin{references}

\reference{AOB04} Antokhin, I.I., Owocki, S.P., \& Brown, J.C. 2004, ApJ, 611, 434
\reference{Chere76} Cherepashchuk, A.M. 1976, SvAL, 2, 138
\reference{HSP04} Henley, D.B., Stevens, I.R., \& Pittard, J.M. 2005, MNRAS, 356, 1308
\reference{GS05} Gardiner, T., \& Stone, J.M. 2005, JCP, 205, 509
\reference{GS06} Gardiner, T., \& Stone, J.M. 2006, JCP, submitted
\reference{GOC97} Gayley, K.G., Owocki, S.P., \& Cranmer, S.R. 1997, ApJ, 475, 786
\reference{LMM90} Luo, D., McCray, R., \& Mac Low, M.-M. 1990, ApJ, 362, 267
\reference{MTD99} Monnier, J.D., Tuthill, P.G., \& Danchi, W.C. 1999, ApJ, 525, L97
\reference{Pittard98} Pittard, J.M. 1998, MNRAS, 300, 479
\reference{Pollock87} Pollock, A.M.T. 1987, ApJ, 320, 283
\reference{PU76} Prilutskii, O. \& Usov, V. 1976, SvA, 20, 2
\reference{Quirk94} Quirk, J.J. 1994, IJNMF, 18, 555
\reference{Rauw05} Rauw, G., et al. 2005, A\&A, 432, 985
\reference{Roe81} Roe, P.L. 1981, JCP, 43, 257
\reference{RL79} Rybicki, G.B., \& Lightman, A.P. 1979, Radiative Processes in Astrophysics (New York, NY: John Wiley \& Sons)
\reference{SMD98} Sanders, R., Morano, E., \& Druguet, M.C. 1998, JCP, 145, 511
\reference{SBP92} Stevens, I.R., Blondin, J.M., \& Pollock, A.M.T. 1992, ApJ, 386, 265
\reference{SP94} Stevens, I.R., \& Pollock, A.M.T. 1994, MNRAS, 269, 226
\reference{TMD99} Tuthill, P.G., Monnier, J.D., \& Danchi, W.C. 1999, Natur, 398, 487
\reference{vdH01} van der Hucht, K.A. 2001, NewAR, 45, 135
\reference{Walder98} Walder, R. 1998, Ap\&SS, 260, 243

\end{references}
\end{document}